\documentclass[sigplan,screen,nonacm]{acmart}

\setcopyright{none}

\usepackage[most]{tcolorbox}
\usepackage{enumitem}
\usepackage{listings}
\usepackage{cuted} 
\usepackage{booktabs, tabularx, threeparttable, makecell}
\usepackage[table]{xcolor}
\definecolor{openframe}{HTML}{E38B2E}
\definecolor{problemneed}{HTML}{6BAA44}
\definecolor{productintro}{HTML}{4B88CF}
\definecolor{persuasive}{HTML}{7052A3}
\definecolor{closure}{HTML}{A24667}
\definecolor{others}{HTML}{6B7A89}

\usepackage{graphicx}
\usepackage{caption}
\usepackage{booktabs, tabularx, threeparttable}
\usepackage{svg} 
\svgpath{{Figures/}} 
\usepackage{etoolbox}

\newtcolorbox{promptblock}[1][]{
  enhanced, breakable,
  colback=darkred!4!white,      
  colframe=darkred!80!black,    
  colbacktitle=darkred!90!black,
  coltitle=white,               
  boxrule=0.5pt,
  arc=1.5mm,
  left=6pt, right=6pt, top=4pt, bottom=4pt,
  boxsep=2pt,
  fonttitle=\bfseries,
  fontupper=\small,
  title=#1
}

\setcopyright{acmlicensed}
\copyrightyear{2026}
\acmYear{2026}
\acmDOI{XXXXXXX.XXXXXXX}
\acmConference[WWW ’26 Companion]{The Web Conference 2026 Companion}{April 13–17, 2026}{Dubai, UAE}

\acmBooktitle{WWW '26: The ACM Web Conference 2026, 
June 03--05, 2026, Dubai, United Arab Emirates}

\makeatletter
\patchcmd{\maketitle}{\@title@typeset}{\@title@typeset\vspace{-0.6\baselineskip}}{}{}

\patchcmd{\@abstract}{\vskip\baselineskip}{\vskip0.35\baselineskip}{}{}
\makeatother

\begin{document}

\title{MLLM-VADStory: Domain Knowledge-Driven Multimodal LLMs for Video Ad Storyline Insights}


\author{Jasmine Yang}
\affiliation{%
  \institution{Meta}
  \city{New York}
  \state{NY}
  \country{USA}}
\email{yumingyang@meta.com}

\author{Poppy Zhang}
\affiliation{%
  \institution{Meta}
    \city{New York}
  \state{NY}
  \country{USA}}
\email{poppyzhang@meta.com}

\author{Shawndra Hill}
\affiliation{%
  \institution{Meta}
  \city{New York}
  \state{NY}
  \country{USA}}
\email{shawndrahill@meta.com}

\renewcommand{\shortauthors}{Yang et al. (2026)}


\begin{abstract}
  We propose MLLM-VADStory, a novel domain knowledge-guided
  multimodal large language models (MLLM) framework to systematically quantify and generate insights for video ad storyline understanding at scale. The framework is centered on the core idea that ad narratives are structured by functional intent, with each scene unit performing a distinct communicative function, delivering
  product and brand-oriented information within seconds. MLLM-VADStory segments ads into functional units, classifies each unit's functionality using a novel advertising-specific functional role taxonomy, and then aggregates functional sequences across ads to recover data-driven storyline structures. Applying the framework to 50k social media video ads across four industry subverticals, we find that story-based creatives improve video retention, and we recommend top-performing story arcs to guide advertisers in creative design. Our framework demonstrates the value of using domain knowledge to guide MLLMs in generating scalable insights for video ad storylines, making it a versatile tool for understanding video creatives in general.
\end{abstract}



\maketitle

\section{Introduction}
The narrative structure in video advertisements (``creatives'') sits along a continuum \cite{escalas2003advertising}, from non-narrative ads which usually feature visual mashups focusing on product promotions and features to narrative ads that embed stories around product experiences or personal transformation through product use. While stories have been shown as powerful vehicles for entertaining consumers in different media \citep[e.g.][]{toubia2021quantifying,shachar2025sell,hartmann2018super,knight2025building}, there is little understanding of how stories drive ad performance and, most importantly, what makes certain ad stories more effective than others. Understanding this design space can help advertisers and platforms seeking actionable insights at scale optimize their creative strategies. 

This paper introduces \textit{MLLM-VADStory}, a MLLM-based framework driven by advertising domain expertise, to systematically quantify and measure video ad stories and their structure arcs at scale. In contrast to prior work on structuring stories through emotions and information flow \cite{toubia2021quantifying,knight2025building}, our approach is built on a novel idea that storytelling in short-form video creatives is \textit{functionally} different from traditional story structures (e.g., Freytag's pyramid \citep[e.g.][]{yang2021design}): ad stories must deliver their message within seconds, blending product and brand-oriented information to achieve specific marketing goals. Thus, ad stories can be viewed as a structured sequence of functional intents, with each scene unit serving a specific communicative purpose. 

Our framework integrates visual and audio signals to decompose a video into semantically coherent functional units, and then guides MLLMs to characterize the functional intent of each unit based on a novel functional role taxonomy we built grounded in advertising domain expertise. Summarizing functional sequences across video creatives reveals data-driven storyline structures that can be linked to ad performance metrics through modeling to generate insights at scale. 

We apply this framework to 50k active video ads across four industry subverticals to derive best practices for creative story design. We show that stories are effective at improving video retention and the best-performing story structure arcs vary by industry subverticals. Analysis across ads reveal early patterns that using hook-first and problem-driven arcs may help improve dwell rates, while showing social proof and clear solutions may help improve product conversions. 

Our contributions are threefold: 
\begin{itemize}
    \item To the best of our knowledge, this is the very first work that leverages MLLMs for video creative storyline understanding using large-scale ads creatives, given the scalability and computational complexity of analyzing video creatives.   
    \item We propose \textit{MLLM-VADStory}, a novel MLLM-based video story framework grounded in advertising domain expertise, to capture data-driven storyline structures through developing a unique functional role taxonomy and structuring storylines in video ads by their underlying functional intent in communication. 
    \item We derive interpretable and actionable story insights at scale and demonstrate the value of our framework for video creative understanding, making it a versatile tool for video insight generation. 
\end{itemize}

\section{Methodology: MLLM-VADStory}
\begin{figure}
    \centering
    \includegraphics[width=1.1\linewidth]{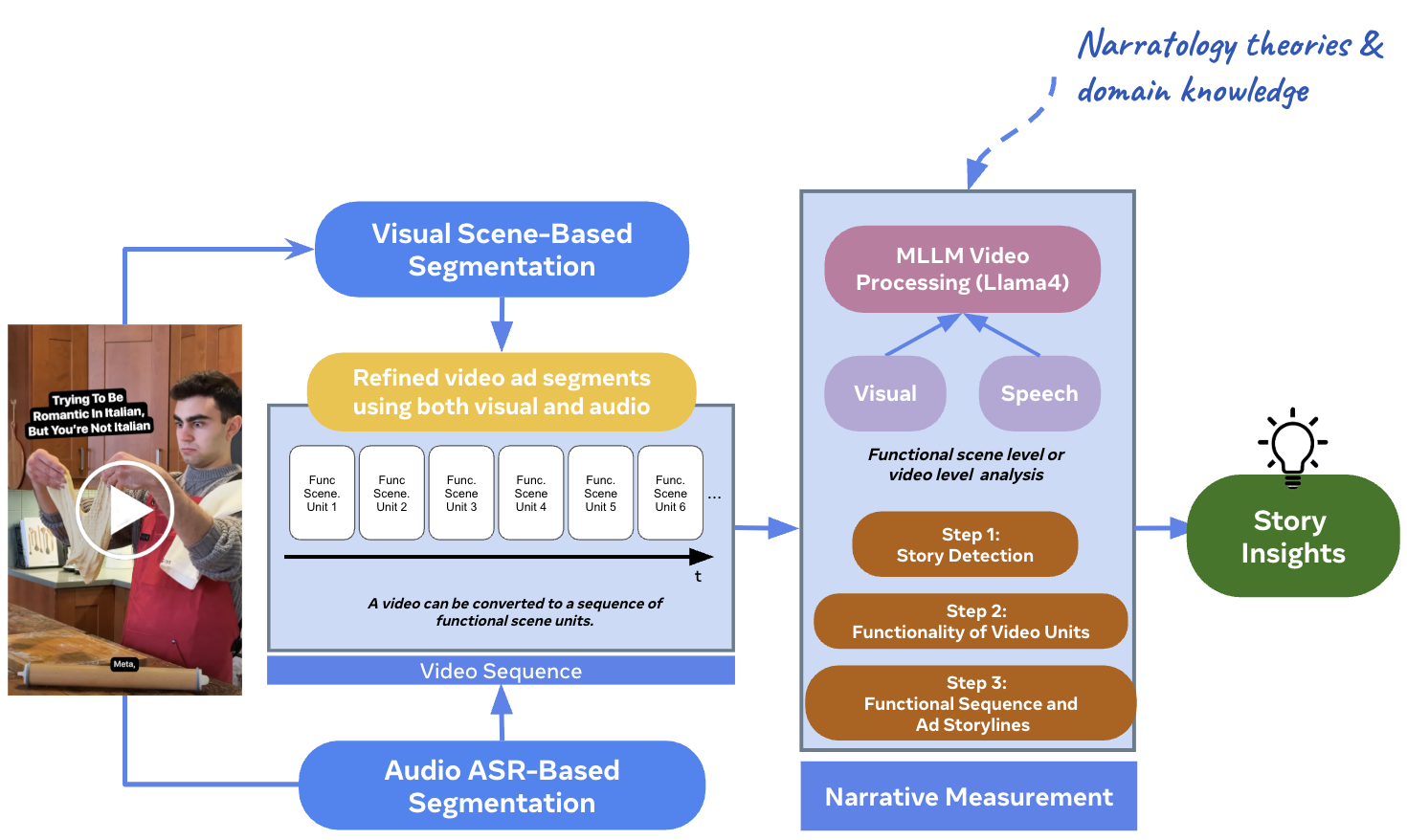}
    \caption{Video Analysis Pipeline for Story Understanding}
    \label{fig:video-pipeline}
\end{figure}

We propose, \textit{MLLM-VADStory}, a domain knowledge-driven framework that integrates visual and audio modalities, leveraging MLLMs to systematically quantify storyline structures embedded in ad creatives. Figure \ref{fig:video-pipeline} illustrates the framework. The framework begins by using visual and speech-based signals to decompose a video into semantically coherent segments, corresponding to functional shifts in a narrative (e.g., problem framing, product reveal, or feature explanation). Then, we use a 3-step approach for narrative understanding: In step 1, we pass visual and speech cues across units to detect whether there is a story in the video creative using MLLM (in our case, Llama MLLM Video Understanding Model\footnote{\url{https://www.llama.com/}}), based on a story definition well-supported by domain knowledge. In step 2, each individual unit is processed using Llama MLLM that jointly reasons over visual and textual cues to generate functional role predictions, guided by narratology principles and advertising domain expertise. Finally, in step 3, we summarize detected narrative functional sequences to obtain scalable data-driven story structures. We link stories and story structures with ad performance metrics to obtain scalable insights using ML models for predictive analysis (see Section \ref{section-application}).

\subsection{Video Segmentation into Functional Units}
We propose a content-aware multimodal segmentation approach that integrates both visual and audio-linguistic cues. Instead of relying solely on visual-based segmentation (PySceneDetect’s content-aware detector\footnote{\url{https://www.scenedetect.com/docs/latest/api/detectors.html}} with an adaptive threshold) which may over-segment a video creative if it consists of visual mashups with scenes serving the same functions, we also extract natural speech boundaries using Whisper\footnote{\url{https://github.com/openai/whisper}}, a robust speech recognition model. We derive speech segmentation timestamps from both temporal cues (e.g., pauses, silence) and linguistic transitions (e.g., conjunctions such as ``and so'' and ``and then''). The two segmentation streams are then synchronized and refined to produce functional units that capture how narrative transitions occur in video ad creatives.

\begin{table*}[t]
\centering
\scriptsize
\renewcommand{\arraystretch}{1.3}
\setlength{\tabcolsep}{1pt}
\begin{threeparttable}
\caption{Functional Role Taxonomy for Video Creatives}
\label{tab:functional-role-taxonomy}
\begin{tabular}{>{\raggedright\arraybackslash}m{5.0cm} >{\raggedright\arraybackslash}m{13cm}}
\toprule
\rowcolor{openframe!20}
\textbf{Opening} & 
\textbf{Hook:} Grabs viewers' attention or interest but not always related to products; appears at the first few seconds of a video.
\textbf{Establish Context:} Sets up the status quo—who, where, or when—of the story before story progression. \\
\midrule
\rowcolor{problemneed!20}
\textbf{Problem, Need \& Challenge} &
\textbf{Problem Setup:} Presents a problem, need, or pain point to resolve for the first time 
\textbf{Problem Agitation:} Amplifies the problem to make it relatable or severe. \\
\midrule
\rowcolor{productintro!20}
\textbf{Product Introduction \& Explanation} &
\textbf{Feature Explanation:} Explains product features and why it delivers benefits; goes beyond just showing.  
\textbf{Product Highlight:} Presents key product attributes or benefits (surface-level showcasing, not deep explanation). 
\textbf{Demonstration/Trial:} Shows the product being used or tested to accomplish a task
\textbf{Comparison:} Contrasts product with competitors or previous states. 
\textbf{Social Proof:} Shows crowd reviews or testimonials from other people. 
\textbf{Solution Reveal:} Presents product as solution to a problem \\
\midrule
\rowcolor{persuasive!20}
\textbf{Persuasive Framing} &
\textbf{Emotional Appeal:} Uses emotions to connect with and engage the audience. 
\textbf{Humor:} Uses comedic elements to entertain and make the message more relatable. 
\textbf{Aspirational Vision:} Depicts an ideal lifestyle or future enabled by the product. 
\textbf{Promotion:} Communicates offer mechanics: discount, bundle, code, pricing terms or value prop specifics (without urgency language). 
\textbf{Urgency Trigger:} Adds time pressure to accelerate action. 
\textbf{Scarcity Trigger:} Highlights limited availability to create FOMO and action; distinct from generic promotion. 
\textbf{Warning:} Alerts viewers to potential risks or disclaimers. \\
\midrule
\rowcolor{closure!20}
\textbf{Closure \& Identity} &
\textbf{Call-to-Action:} Cues to act; drive immediate action. 
\textbf{Outcome:} Shows post-intervention payoff or transformation. 
\textbf{Branding Moment:} Displays brand identity (e.g., logo, tagline, slogans).
\textbf{Insight/Philosophy:} Expresses brand or product philosophy; leads viewers to discover something new about life, human behavior or how the world works. 
\textbf{Teaser/Cliffhanger:} Leaves the narrative open-ended to invite re-engagement. \\
\midrule
\rowcolor{others!20}
\textbf{Others} &
\textbf{Visual Filler:} Provides transitional pacing without narrative contribution. \\
\bottomrule
\end{tabular}
\end{threeparttable}
\end{table*}

\subsection{Narrative Measurement}
\subsubsection{(S1) Story Detection} We first detect whether a video creative contains stories by guiding Llama MLLM to evaluate ads based on both its visual scenes and speech transcript across all units according to domain knowledge-driven story definition. Supported by past work on narratives\citep[e.g.][] {piper2021narrative,genette1980narrative,mckee1997substance,van2019happens}, we define a story as ``an account of an event or a sequence of connected events that leads to a transition from an initial state to a later stage or outcome''. 


\subsubsection{(S2) Functionality of Video Units} 

\noindent \paragraph{Functional Role Taxonomy.} To understand how stories unfold through the sequence of functionality, we then propose a granular, non-overlapping, and flexible functional role taxonomy (see Table \ref{tab:functional-role-taxonomy}) grounded in advertising domain knowledge to characterize the functionality each unit serves. The taxonomy covers six core categories that collectively span the full video spectrum.  \noindent (1) \textbf{\textcolor{openframe}{Opening and Framing}}: Captures viewers attention to the ad at the start and establishes the setting. 
\noindent (2) \textbf{\textcolor{problemneed}{Problem, Need, and Challenge}}: Presents the tension, need, or pain point that motivates the narrative. 
\noindent (3) \textbf{\textcolor{productintro}{Product Introduction and Explanation}}: Demonstrates how the product or service addresses the problem and communicates its features or benefits. 
\noindent (4) \textbf{\textcolor{persuasive}{Persuasive Framing}}: Highlights the product’s desirability through different cues such as emotions, promotional incentives, time and quality-related cues. 
\noindent (5) \textbf{\textcolor{closure}{Closure and Identity}}: Provides resolution or brand reinforcement by signaling outcomes, calls to action, or brand identity. 
\noindent (6) \textbf{\textcolor{others}{Others}}: Captures transitional or auxiliary elements that maintain pacing or tone without advancing the core narrative.

\begin{figure}[t]
    \centering
    \includegraphics[width=1.1\linewidth]{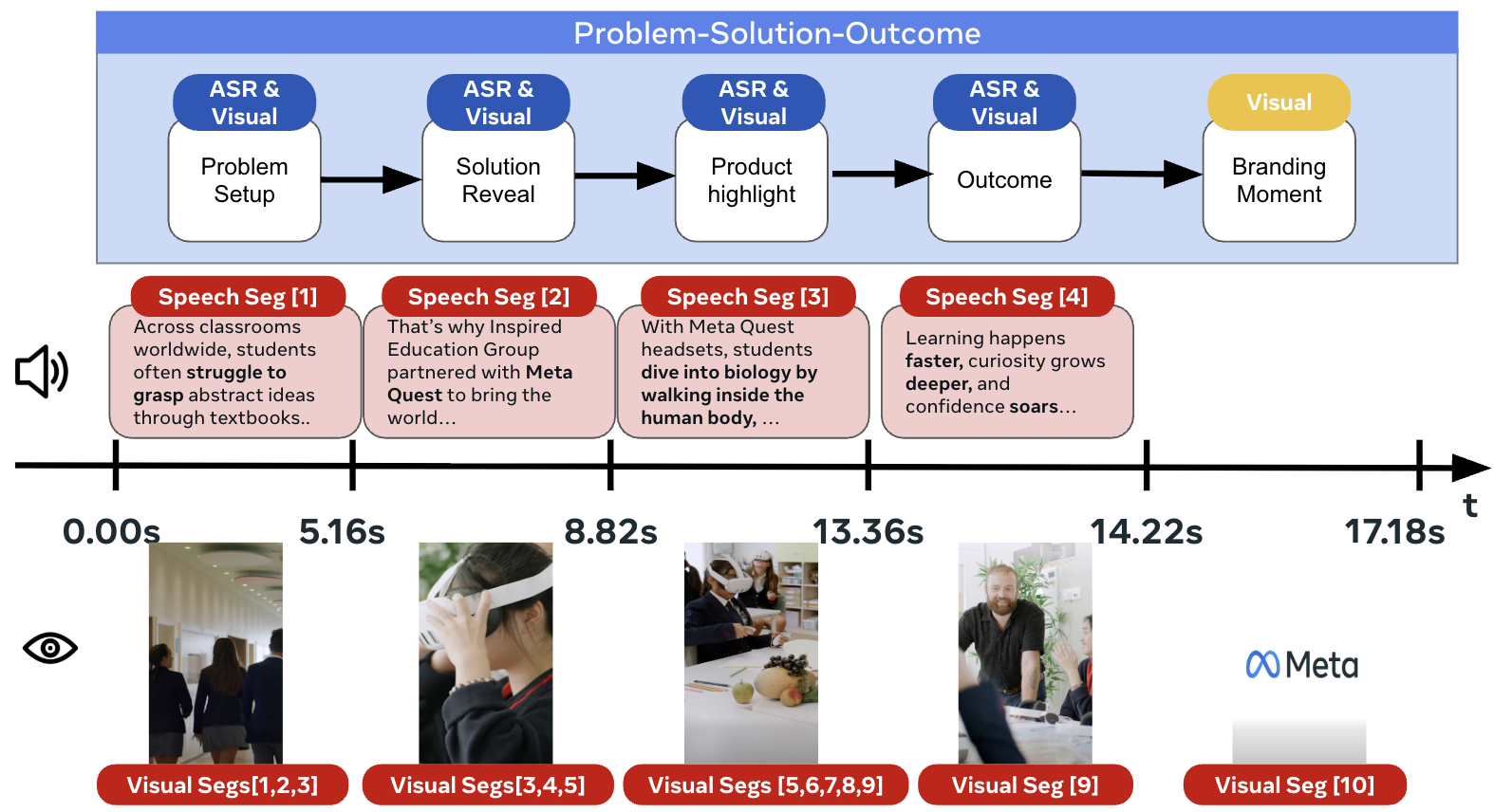}
    \caption{An example storyline for a video creative}
    \label{fig:storyline-example}
\end{figure}

\paragraph{Functional Role Detection.} We use Llama MLLM to automatically infer the primary functionality employed in each functional unit using both visual scenes and speech transcript in the unit, provided the full taxonomy and the descriptions of each role in the prompt. Figure \ref{fig:storyline-example} illustrates an example.

\subsubsection{(S3) Functionality Sequence and Ad Storylines} 
\begin{figure}
    \centering
    \includegraphics[width=\linewidth]{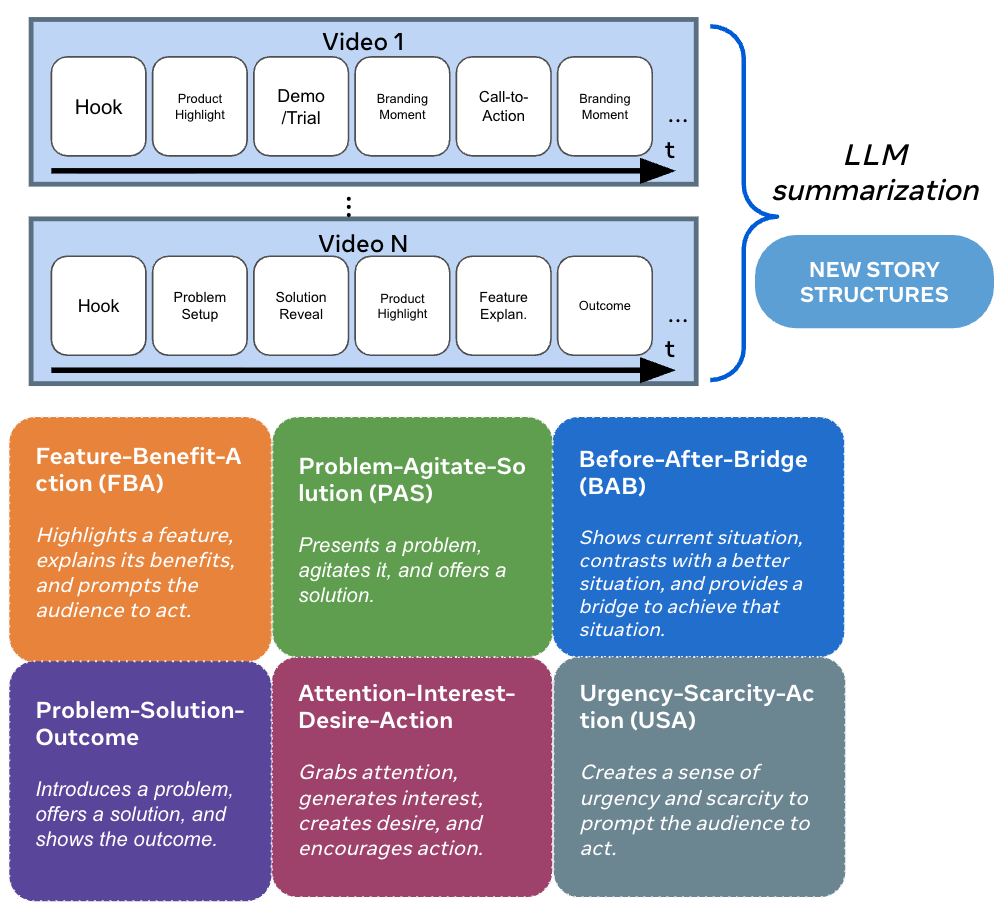}
    \caption{Illustration of Functional Sequence Summarization and Example Storyline Structures}
    \label{fig:llm-summarization}
\end{figure}

We summarize functional sequences across video creatives using Llama‑4-Scout-17B-16E Instruct\footnote{\url{https://ai.meta.com/blog/llama-4-multimodal-intelligence/}} with human-in-the-loop for naming and merging semantically similar structures to obtain macro story structures. Figure \ref{fig:llm-summarization} illustrates the approach and some examples of our data-driven storylines. 

Classifications completed by MLLMs for all steps were manually inspected by creative strategists with domain expertise to ensure accuracy. 

\section{Application}\label{section-application}

\paragraph{Data and Setting} We apply the framework to 50,000 active social media video ads. We focus on English ads with a single video, between 15 seconds and 60 seconds in length, from four industry subverticals (Apparel \& Accessories, Beauty, Food, and Beverages). In addition, we restrict our analysis to ads from U.S. advertisers above a certain minimum daily impressions and budget for the first day since ad launch, to ensure that advertisers invest effort in these creatives and that the results reflect best practices for story design. 

\begin{figure}[t]
  \vspace{-2mm} 
  \centering
  \includegraphics[width=0.8\linewidth]{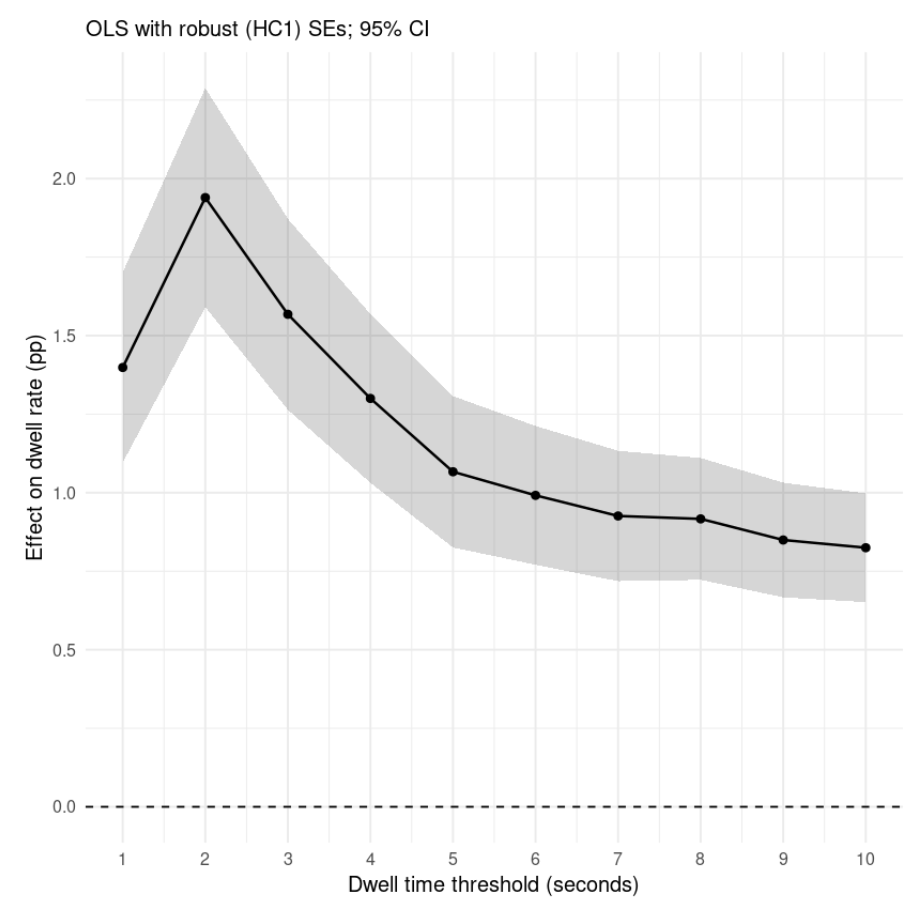}
  \caption{\% Change in Dwell Rate in the First 10 Seconds when Ads Contain Stories}
  \label{fig:perc-lift}
  \vspace{-2mm} 
\end{figure}

\paragraph{Insights for Ad Stories} Across ad creatives, we find that (1) on average 48.2\% of them contain a story, with Beverages (65\%) and Beauty (53.7\%) investing a lot more in storytelling; (2) Larger advertisers invest into stories more, with story concentration decreasing by advertiser size; (3) Stories are used more when campaign objectives chosen by advertisers are for awareness and sales, and (4) Stories are most effective at improving overall video retention in the early seconds. Figure \ref{fig:perc-lift} plots the effect of stories in the first 10 seconds of a video on \% change in average dwell time (defined as \% viewers who watched a certain seconds averaged across videos) using OLS, after controlling for video-related (e.g, has speech, video length), ad and campaign-related features (e.g., campaign objective, audience size). We show that the absolute \% change is highest in 2 seconds. Taking the baseline average dwell rate for non-story ads, this translates to about 5 to 10\% increase in relative dwell rate. 

We also ask MLLM to explain why there is a story in the ad and summarize across creatives to obtain story signals. Agreeing with past marketing insights \cite{shachar2025sell}, we find that dialogues, sharing of personal experiences, inclusion of challenges/conflicts/problem solutions, etc usually signal story existence. On the other hand, the existence of an announcer, use of promotional language, and heavy focus on product features or benefits signal story absence. 


\begin{table}[t]
\centering
\scriptsize
\renewcommand{\arraystretch}{1}
\setlength{\tabcolsep}{3pt}
\caption{Top Storyline Patterns by Ad Metric and Subvertical}
\label{tab:storyline-performance}
\begin{threeparttable}
\begin{tabularx}{0.95\linewidth}{@{}>{\raggedright\arraybackslash}p{0.28\linewidth}l>{\raggedright\arraybackslash}p{0.45\linewidth}@{}}
\toprule
\textbf{Ad Metric by Subverticals} & \textbf{Top-Performing Storylines} & \textbf{Uplift} \\
\midrule
\rowcolor{gray!10}
\multicolumn{2}{@{}l}{\textbf{Dwell 2s Rate}} \\
Apparel \& Accessories & Hook–Feature–Benefit–Action (HFBA) & +5.8\% \\
Beauty & Hook-Problem-Demo-Solution (HPDS) & +4.9\% \\
Food & Problem–Agitate–Solution (PAS) & +6.3\% \\
Beverages & Hook–Problem–Solution (HPS) & +4.1\% \\
\midrule
\rowcolor{gray!10}
\multicolumn{2}{@{}l}{\textbf{Click-Through Rate (CTR)}} \\
Apparel \& Accessories & Attention–Interest–Desire–Action (AIDA) & +8.9\% \\
Beauty & Hook–Feature–Benefit–Action (HFBA) & +2.8\% \\
Food & Attention–Interest–Desire–Action (AIDA) & +4.8\% \\
Beverages & Feature-Benefit-Action (FBA) & +3.9\% \\
\midrule
\rowcolor{gray!10}
\multicolumn{2}{@{}l}{\textbf{Conversion Rate (CVR)}} \\
Apparel \& Accessories & Social–Proof–Action (SPA) & +4.6\% \\
Beauty & Social–Proof–Action (SPA) & +3.7\% \\
Food & Problem–Agitate–Solution (PAS) & +8.5\% \\
Beverages & Feature–Highlight–Explanation–Action (FHEA) & +5.1\% \\
\bottomrule
\end{tabularx}
\end{threeparttable}
\end{table}

\paragraph{Insights for Ad Storyline Structures}. Across subverticals, we find that (1) the most commonly used story structures are Feature-Benefit-Action and Problem-Solution type of arcs (Problem-Agitate-Solution, Hook-Problem-Solution, and Problem-Solution-Outcome); (2) Before-After-Bridge is used a lot more in Beauty followed by Apparel compared to other subverticals; (3) there are more variations in story structure usage in Beauty across videos, compared to other subverticals; and (4) Attention-Interest-Desire-Action appears in the top adoption list for Food, likely due to how appetite and purchase motivation are driven by desire psychologically.

To predict uplift of story structures on ad performance (dwell rate past 2 seconds, CTR, CVR), we fit XGBoost models, which are ML models that leverage gradient-boosted decision trees, to each subvertical while accounting for video traits (e.g., has audio, has speech, video length, aspect ratio) and other metadata (e.g., audience, advantaged targeting). Table \ref{tab:storyline-performance} presents the top performance story structure by the amount they contribute to increases in each ad metric via partial dependence. We find that while the most effective story structure vary by subvertical, there are some trends across the four subverticals. In general, hook-first, problem-driven arcs may help improve dwell rate past 2 seconds. Clean attention to interest/benefit arcs may help improve CTR, and showing social proof and clear solutions may increase CVR. Advertisers can take these insights into account when refining or designing new creatives for further experimentation. 

\section{Conclusion}
We propose MLLM-VADStory, a advertising domain knowledge-driven framework that leverages MLLMs to systematically quantify storyline structures embedded in ad creatives. We apply the framework to video creatives at scale to derive interpretable story insights for advertisers. We hope that this work can inspire more research using domain knowledge to guide MLLMs for video creative understanding. 

\bibliographystyle{ACM-Reference-Format}
\bibliography{references}

@String{Computer = "{IEEE} Computer" }

@incollection{escalas2003advertising,
  title={Advertising narratives: what are they and how do they work?},
  author={Escalas, Jennifer Edson},
  booktitle={Representing consumers},
  pages={283--305},
  year={2003},
  publisher={Routledge}
}

@article{toubia2021quantifying,
  title={How quantifying the shape of stories predicts their success},
  author={Toubia, Olivier and Berger, Jonah and Eliashberg, Jehoshua},
  journal={Proceedings of the National Academy of Sciences},
  volume={118},
  number={26},
  pages={e2011695118},
  year={2021},
  publisher={National Academy of Sciences}
}

@article{shachar2025sell,
  title={Sell Me a Story},
  author={Shachar, Ron and Muchnik, Lev and Netzer, Oded},
  journal={Available at SSRN 5236601},
  year={2025}
}

@inproceedings{piper2021narrative,
  title={Narrative theory for computational narrative understanding},
  author={Piper, Andrew and So, Richard Jean and Bamman, David},
  booktitle={Proceedings of the 2021 Conference on Empirical Methods in Natural Language Processing},
  pages={298--311},
  year={2021}
}

@article{van2019happens,
  title={What happens in Vegas stays on TripAdvisor? A theory and technique to understand narrativity in consumer reviews},
  author={Van Laer, Tom and Edson Escalas, Jennifer and Ludwig, Stephan and Van Den Hende, Ellis A},
  journal={Journal of Consumer Research},
  volume={46},
  number={2},
  pages={267--285},
  year={2019},
  publisher={Oxford University Press}
}

@article{mckee1997substance,
  title={Substance, structure, style, and the principles of screenwriting},
  author={McKee, Robert},
  journal={Alba Editorial},
  year={1997}
}

@book{genette1980narrative,
  title={Narrative discourse: An essay in method},
  author={Genette, G{\'e}rard},
  volume={3},
  year={1980},
  publisher={Cornell University Press}
}

@article{knight2025building,
  title={Building Persuasive Stories with Emotion Sequences},
  author={Knight, Samsun and Liu, Liu and Kornish, Laura J},
  journal={Available at SSRN 5489708},
  year={2025}
}

@article{hartmann2018super,
  title={Super bowl ads},
  author={Hartmann, Wesley R and Klapper, Daniel},
  journal={Marketing Science},
  volume={37},
  number={1},
  pages={78--96},
  year={2018},
  publisher={INFORMS}
}

@article{yang2021design,
  title={A design space for applying the freytag's pyramid structure to data stories},
  author={Yang, Leni and Xu, Xian and Lan, XingYu and Liu, Ziyan and Guo, Shunan and Shi, Yang and Qu, Huamin and Cao, Nan},
  journal={IEEE Transactions on Visualization and Computer Graphics},
  volume={28},
  number={1},
  pages={922--932},
  year={2021},
  publisher={IEEE}
}

\end{document}